\documentclass[preprint,12pt]{elsarticle}

\usepackage{fixltx2e,hyperref}
\usepackage{graphicx,amssymb,here}
\usepackage{hyperref}
\journal{}

\def\1{\mathrm{1}}
\def\2{\mathrm{2}}

\biboptions{sort&compress}
\begin{document}

\begin{frontmatter}

\title{Discovery of nanographene for hydrogen storage solving low reversibility issues
}

\author{Gagus Ketut Sunnardianto$^\mathrm{a,b}$\footnote{Corresponding
author.\\E-mail address : gagu001@brin.go.id   (Gagus Ketut Sunnardianto)}, Yusuf Wicaksono$^\mathrm{c}$, Halimah Harfah$^\mathrm{c}$, Koichi Kusakabe$^\mathrm{a}$}

\address{$^\mathrm{a}$School of Science, Graduate School of Science, University of Hyogo, 3-2-1 Kouto, Kamigori-cho, Ako-gun, Hyogo, 678-1297, Japan
}

\address{$^\mathrm{b}$Research Center for Quantum Physics, National Research and Innovation Agency (BRIN), Kawasan Puspiptek Serpong, Tangerang Selatan, 15314, Indonesia
}

\address{$^\mathrm{c}$Graduate School of Engineering Science, Osaka University, 1-3 Machikaneyama-cho, Toyonaka, Osaka 560-8531, Japan
}

\begin{abstract}
We found good reversibility of hydrogen uptake-release in vacancy-centered hexagonal armchair nanographene (VANG) based on density functional theory calculation. VANG has triply hydrogenated vacancy (V$\textsubscript{111}$) at the center acts as a self-catalytic property to reduce an activation barrier of hydrogen uptake-release. We found remarkable features in an almost equal value of the activation energy barrier of 1.19 eV for hydrogen uptake and 1.25 eV for hydrogen release on V$\textsubscript{111}$ of VANG. The dehydrogenation showed slightly exothermic and the hydrogenation became slightly endothermic, suggesting the efficiency of hydrogen uptake-release. In high hydrogen coverage, the quintuply hydrogenated vacancy (V$\textsubscript{221}$) is formed with some hydrogenated located in the in-plane and armchair edges. This structure produces an exothermic hydrogen release from the in-plane with an energy barrier of not more than 2 eV. This finding potentially addresses the low reversibility issues in the organic chemical hydrides as hydrogen storage materials.

\end{abstract}

\end{frontmatter}
\section{Introduction}

The rapid increase of world energy consumption has created considerable demands in various energy resources \cite{Konstantinos,Wiley}. Fossil fuels will become more limited than in the past century due to CO$\textsubscript{2}$ emission; thus, clean alternative energy sources are requested to be developed. Among them, hydrogen has been considered as one of the most promising clean alternatives \cite{ADEM:ADEM200600018,Louis,Satyapal2007246,ANIE:ANIE200806293}. However, there are several remaining barriers against the promotion of hydrogen society for various applications \cite{Louis}. The problems include those in the hydrogen supply chain, where organic chemical hydrides (OCH), which are expected to be substitutes of unsafe compressed hydrogen gas or liquid forms in transportation show an uneasy hydrogen release process, preventing technology using OCH away from wide acceptance \cite{Endo_OH}.

We have recently reported that a hydrogenated atomic vacancy of graphene may solve the problem \cite{gagus2,gagus3}. A key atomic structure is the triply hydrogenated atomic vacancy (V$\textsubscript{111}$) of graphene observed in nature \cite{Ziatdinov2014}.
Using the hydrogen uptake-release reaction process occurring at V$\textsubscript{111}$, we can solve the energy barrier problem existing in hydrogen technology using graphene \cite{PhysRevB.84.041402}. The pristine graphene surface is too inert with a large energy barrier expected to be from 2.7 to 3.3 eV to allow tractable hydrogen dissociative chemisorption \cite{C2CP42538F,:/content/aip/journal/jap/93/6/10.1063/1.1555701,ao}. By realizing ideal chemical reaction paths for hydrogen uptake-release processes with only the 1.3 eV-level barriers, we may start realistic use of graphene for hydrogen storage application, which had been a long-term issue of graphene technology \cite{PhysRevB.84.041402,C2CP42538F,Patchkovskii26072005,PhysRevB.77.035427,Elias610,C0EE00295J,PhysRevB.85.155408,Zhou2012245}.

Furthermore, although graphene \cite{RevModPhys.81.109,Novoselov2004, Novoselov2005, Geim2007} is outstanding for various applications, its planar atomic layer form is not always optimal. The best handling might be found in a molecular form for chemical reactivity. Therefore, we should search for good candidates among various molecular structures of nanographene \cite{Cai2010,C5SC03280F,doi:10.1021/nn100971s,doi:10.1021/jacs.6b04092}. 

In our recent study, we obtained a stable molecular structure of nanographene, with a center-atomic vacancy of V$\textsubscript{111}$. 
The new form of nanographene is called vacancy-centered hexagonal armchair nanographene (VANG) \cite{mori}. 
The structural stability of VANG molecules is successfully tested using density-functional-theory simulations. 
The VANG molecule has an interesting electronic property, where two degenerate zero modes appear in the electronic states. 
The degeneracy is topologically protected as far as a mirror symmetry axis is maintained, even in the deformation process \cite{doi:10.7566/JPSJ.86.034802}. 
These electronic characteristics suggest the hopeful plausibility of VANG even for important chemical reactivity for hydrogen storage applications. 

In this study, we conducted a comprehensive study to identify the hydrogen uptake through V$\textsubscript{111}$ of VANG, hydrogen migration pathways from vacancy to in-plane surface and armchair edges of VANG, hydrogen release from in-plane surfaces in larger amounts of hydrogen stored at in-plane and armchair edges of VANG. It is worth noting that the observation on the reaction pathways, energy barrier, exothermic or endothermic reaction, and different energy between reactants and product are essential for the hydrogen uptake-release process. 

\section{Material and Methods}

\begin{figure}[H]
\centering
\includegraphics[width=6cm]{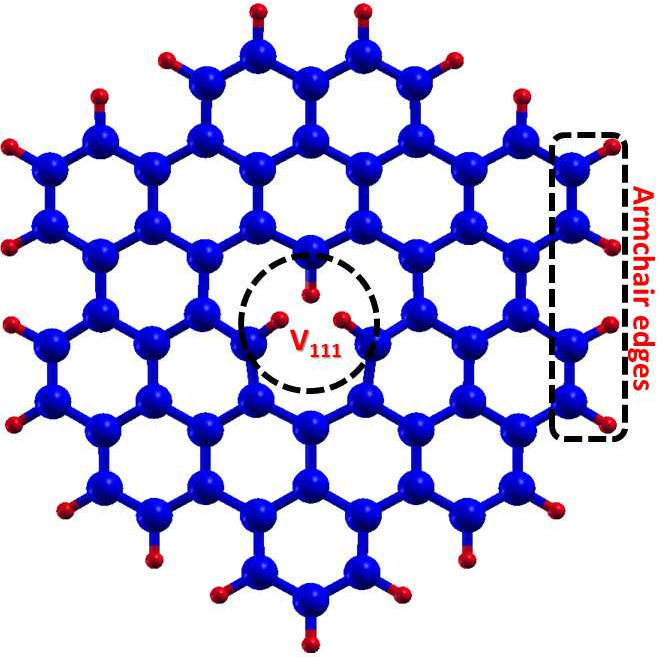}
\caption{Structure of a VANG molecule. 
Triply hydrogenated vacancy (V$\textsubscript{111}$) is located at the center of VANG, and the periphery of VANG consists of armchair edges. 
The VANG molecule has a deformed non-planar structure because of V$\textsubscript{111}$ at the center of the molecule. The blue and red atoms denote carbon and hydrogen, respectively.}
\label{vang}
\end{figure}

A nanographene molecule model used for the simulation is VANG. This nanographene molecule consists of 60 carbon atoms and 24 hydrogen atoms (C$\textsubscript{60}$H$\textsubscript{24}$) \cite{mori}. It consists of V$\textsubscript{111}$ at the center \cite{Ziatdinov2014} and armchair edges at the periphery. Here, each carbon atom at the periphery is mono-hydrogenated, as shown in Figure \ref{vang} using XCrySDen \cite{Kokalj2003}. In our recent theoretical study, we showed the stability of the VANG nanographene in a DFT simulation \cite{mori}. 

The calculations were conducted by using DFT calculation \cite{martin,Hohenberg1964} implemented in the {\sc Quantum ESPRESSO} code \cite{QE-2009}. 
The local density approximation and ultra-soft pseudopotential were adopted \cite{Kohn1965, Perdew1992,Vanderbilt1990, Rappe1990}. 
The nudged-elastic-band method was used to determine the minimum energy pathway of a reaction path and energy barrier for dissociative-chemisorption, migration and desorption processes \cite{PhysRevLett.72.1124,:/content/aip/journal/jcp/113/22/10.1063/1.1323224}. The difference between the total energy of the system at the initial state and the highest energy for the transition state defined as the value of the energy barrier were investigated.

The parameters for hydrogenation of VANG in a super-cell are the energy cutoff of 24 Ry and 240 Ry for the plane wave expansion of the wave function and augmented charge, respectively. The distance between nanographene planes was separated by 10 \r{A} to avoid interaction between layers. 
The convergence criterion for the structural optimization was that the total absolute value of the interatomic force vector became less than $10^{-5}Ry/a.u$.

\section{Results and Discussion}
\subsection{Reaction pathways of H$\textsubscript{2}$ dissociative-chemisorption and desorption of H atoms on the VANG surface}

\begin{figure}[H]
\centering
\includegraphics[width=13cm]{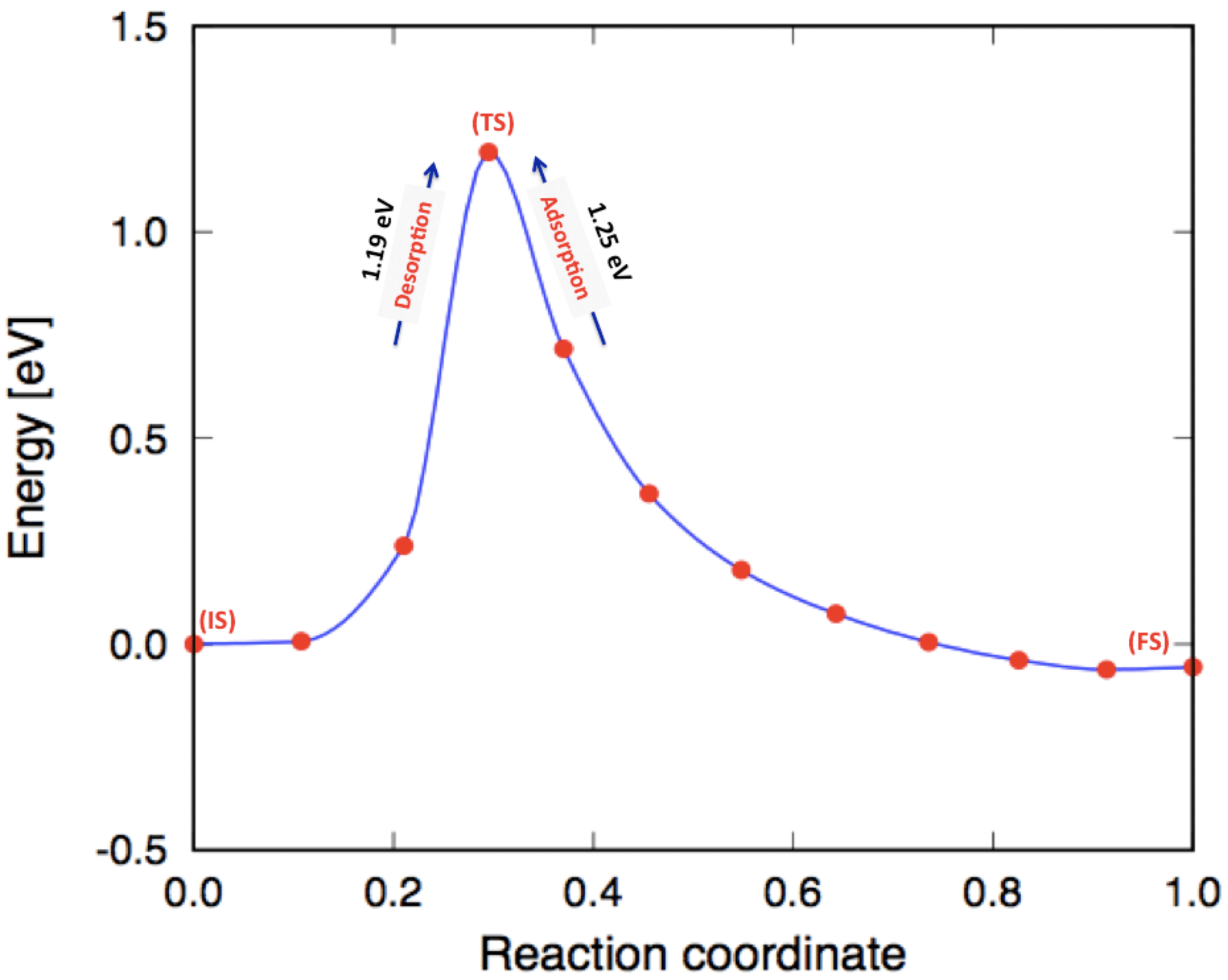}
\caption{A minimum energy pathway of desorption of hydrogen atoms and dissociative-chemisorption of H$\textsubscript{2}$  on VANG. The desorption reaction is from left to right, whereas the hydrogenation reaction is from right to left. The blue and red atoms denote carbon and hydrogen, respectively. IS, TS, and FS represent initial, transition, and final states, respectively.}
\label{VANG_reaction1}
\end{figure}

In our previous study, we successfully observed the hydrogenation and dehydrogenation of graphene vacancy \cite{gagus3}, which has interesting reaction characteristics. 
Since a similar reaction was expected in a molecular nanographene, 
we also examined the hydrogenation and dehydrogenation of VANG. Figure \ref{VANG_reaction1} shows the obtained potential energy surface. An interesting feature is that the hydrogenation process is slightly endothermic; whereas the dehydrogenation process is slightly exothermic. 
This property with reversed stability in the redox reaction would be much favorable for hydrogen storage use. 

\begin{figure}[H]
\centering
\includegraphics[width=13cm]{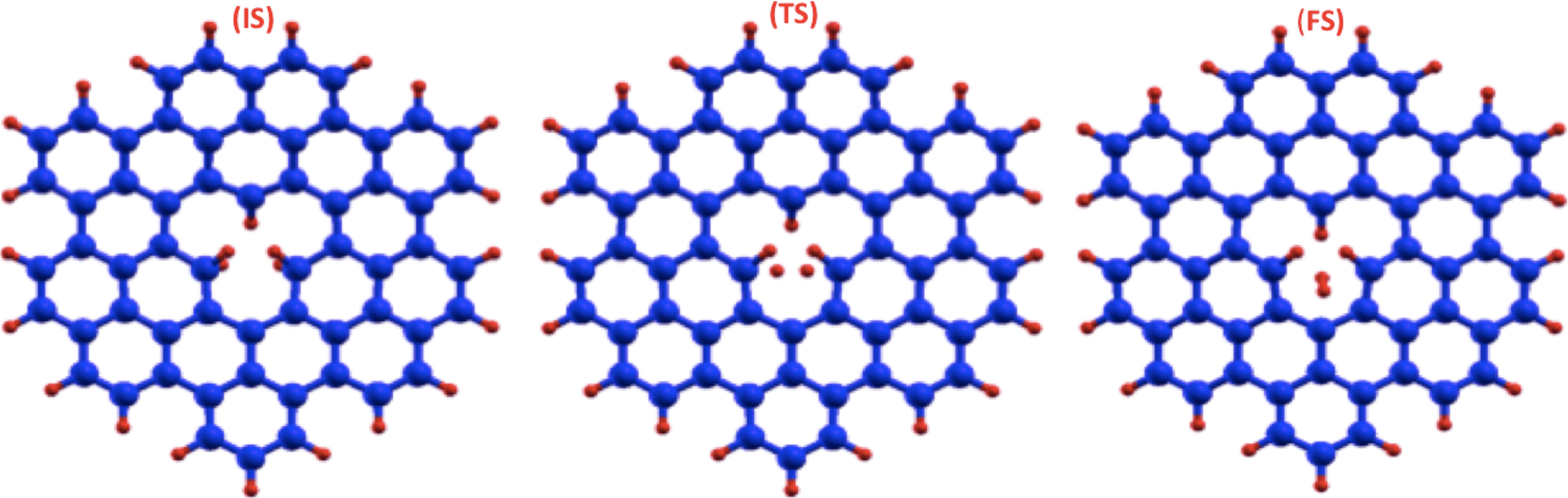}
\caption{Structures appearing along the pathway of Figure \ref{VANG_reaction1} . Left: initial state (IS); center: transition state (TS); right: final structure (FS). The blue and red atoms denote carbon and hydrogen, respectively.}
\label{VANG_reaction1T}
\end{figure}

If we compare the VANG (C$_{60}$H$_{24}$) with V$\textsubscript{111}$ of graphene \cite{gagus3}, the structure of VANG is rather deformed from a planar form. 
This is because of its center hydrogenated vacancy. We identified a buckled structure of VANG, which depends on the orientation of the adsorbed hydrogen atoms. If a hydrogen atom is adsorbed at the top of a carbon atom, the hydrogenated carbon atom buckles out of the plane (above the graphene plane) because of the formation of local $sp^{\mathrm{3}}$-like bondings, whereas if hydrogen atom is adsorbed in the below of carbon atom, the hydrogenated carbon atom buckles out of the plane (below the graphene plane).
The deformation is much apparent when hydrogen atom is adsorbed on VANG 
to create hydrogenated VANG. In other words, the molecular structure of VANG changes its geometrical form along the hydrogenation path. 

The deformation effect of VANG causes a qualitative difference in the potential energy surface between VANG and hydrogenated vacancy of graphene \cite{gagus3}. There appear the dehydrogenation of VANG showing slightly exothermic and the hydrogenation showing slightly endothermic. These characteristics are caused by easier deformation in the VANG structure than graphene with mono-vacancy. 

In our previous paper \cite{gagus1}, we observed the charge-transfer rates (CTR) of a hydrogen atom at an on-top site in deformed graphene surfaces. 
In convex and concave graphene surfaces, CTR decreases their values from the planar pristine graphene. 
This result suggests that the desorption process in a deformed structure becomes easier than in a planar structure, because decreased CTR indicates weakened carbon-hydrogen bonding. 

Similarly, the deformed structure in the hydrogenated VANG makes adsorbed hydrogen atoms slightly unstable, resulting in the easiness of the dehydrogenation process. 
Thus, we found the dehydrogenation of the hydrogen atom from VANG showing a little bit of exothermic nature. 
The important finding that the dehydrogenation process is slightly exothermic may solve the current existing problem of endothermic dehydrogenation processes with large energy barrier height appearing in graphane and Methylcyclohexane (MCH).

\subsection{Migration of a hydrogen atom from a vacancy site (V$\textsubscript{221}$) to an in-plane surface and armchair edges of VANG}

\begin{figure}[H]
\centering
\includegraphics[width=13cm]{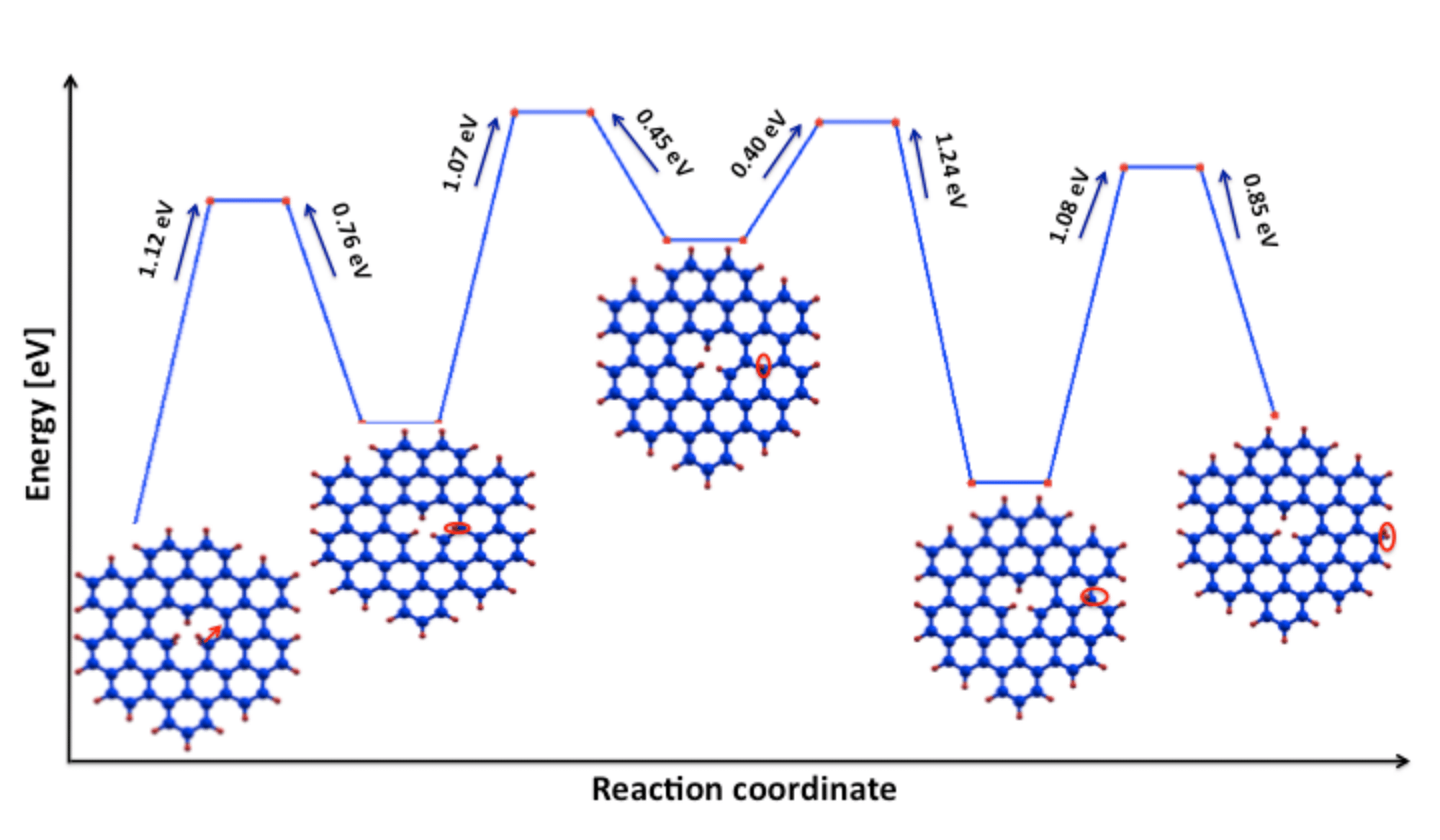}
\caption{Migration pathway of a hydrogen atom starting from a vacancy site (V$\textsubscript{221}$) to an in-plane surface and armchair edges of VANG. The blue and red atoms denote carbon and hydrogen, respectively.
}
\label{VANG_reaction3}
\end{figure}

The Migration of a hydrogen atom from the densely hydrogenated vacancy of VANG (V$\textsubscript{221}$) to an in-plane surface of VANG could occur in real nature. The migration motion may be enhanced by increasing temperatures in a real experiment. In this possible migration path of a hydrogen atom, we calculated the energy barrier of the migration process of a hydrogen atom from the vacancy to the in-plane surface and armchair edges of VANG, as shown in Figure \ref{VANG_reaction3}. At the end of this reaction, we obtained the hydrogenated carbon sites at the vacancy (V$\textsubscript{211}$) and armchair edges of VANG. 

Figure~\ref{VANG_reaction3} shows four stepwise reaction pathways required to migrate a hydrogen atom from the vacancy to edge sites.
The initial state is VANG, with five hydrogen atoms adsorbed at the vacancy site (V$\textsubscript{221}$). The final state is VANG, with four hydrogen atoms adsorbed at the vacancy site (V$\textsubscript{211}$) and one hydrogen atom adsorbed at armchair edges. At the beginning of the reaction, one hydrogen atom in a dihydrogenated carbon vacancy site migrates to the nearest neighbor bare carbon atom through a transition state, with an energy barrier of about 1.12 eV. Furthermore, the migrated hydrogen continually moves to the next-nearest neighbor site of a bare carbon atom, passing through the second transition state with an energy barrier of 1.07 eV. Then, in the third step, the migrated hydrogen continually moves to the next-nearest neighbor site of a bare carbon atom near armchair edges (hop on a more favorable site), overcoming a very small barrier of 0.4 eV. Finally, the adsorbed migrated hydrogen atoms reach the armchair edge with relative ease. There are four steps in this migration path to reach the final state, which has the total energy higher than the initial state. Therefore, the migration path of one hydrogen atom from the vacancy site to an in-plane site and armchair edges shows an endothermic reaction with a relatively small energy barrier of less than 1.25 eV. Subsequently, we expect a hydrogen atom at another dihydrogenated vacancy site also moves to the in-plane site. As a result, V$\textsubscript{111}$ is produced due to the migration of hydrogen atom from the dihydrogenated site to the in-plane surface. Thus, successive hydrogen storing is generated again through V$\textsubscript{111}$.

\subsection{Hydrogen desorption process of nanographene with high-hydrogen density 
}

Further enhanced hydrogenation in larger amounts of hydrogen adsorption is possible in the experiment, when graphene is exposed to large doses of hydrogen atoms. During subsequent continuous hydrogenation, a larger amount of hydrogen can be adsorbed; thus, it has the possibility for densely charging hydrogen in VANG. Therefore, it is important to investigate the reaction pathway of hydrogen uptake-release under a high concentration of hydrogen in VANG. 

\begin{figure}[H]
\centering
\includegraphics[width=13cm]{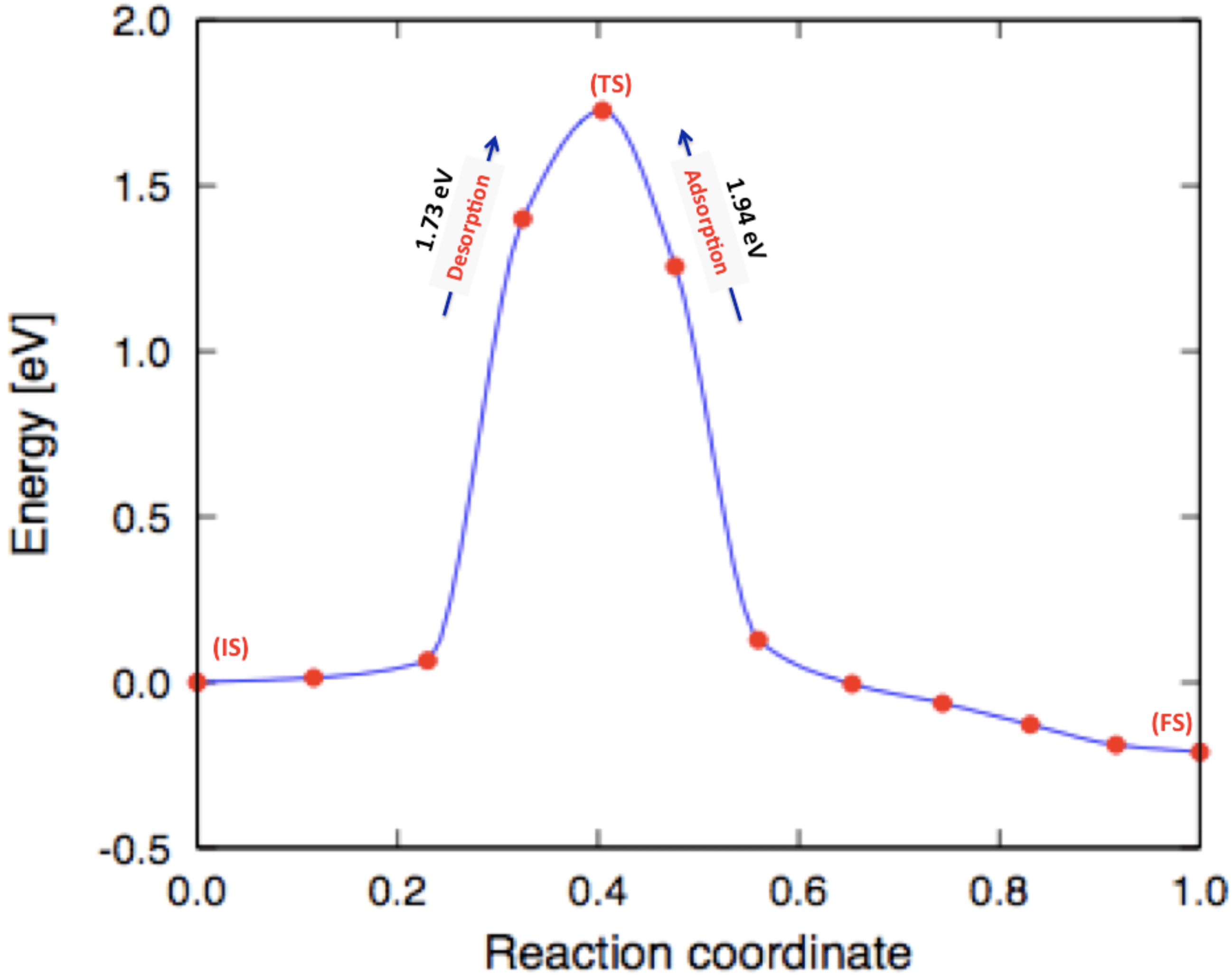}
\caption{Energy profiles for the reactions of the desorption barrier of two hydrogen atoms from in-plane and its reverse reaction. The Blue and red colors indicate carbon and hydrogen atoms, respectively. IS, TS, and FS represent initial, transition, and final states, respectively.}
\label{dua}
\end{figure}

The results are presented in Figure~\ref{dua}, showing the energy pathway of the desorption process of the two hydrogen atoms from the in-plane surface. As the reaction proceeds, the hydrogen bond weakens and passes through the transition state barrier to obtain the final states, where two hydrogen atoms desorb, forming hydrogen molecules. The reaction pathway for the desorption barrier of two hydrogen atoms adsorbed on the surface proceeds through the transition state with an energy barrier of 1.73 eV, forming a hydrogen molecule located above the plane. Reversely, the hydrogen molecule dissociation-chemisorption attempting to cross faces an energy barrier of 1.94 eV prior to the in-top adsorption structure is formed. The activation barrier of releasing hydrogen is not more than 2 eV, indicating not required higher energy to release hydrogen atoms from the in-plane surface. More importantly, the releasing process is a non-endothermic reaction.

\begin{figure}[H]
\centering
\includegraphics[width=13cm]{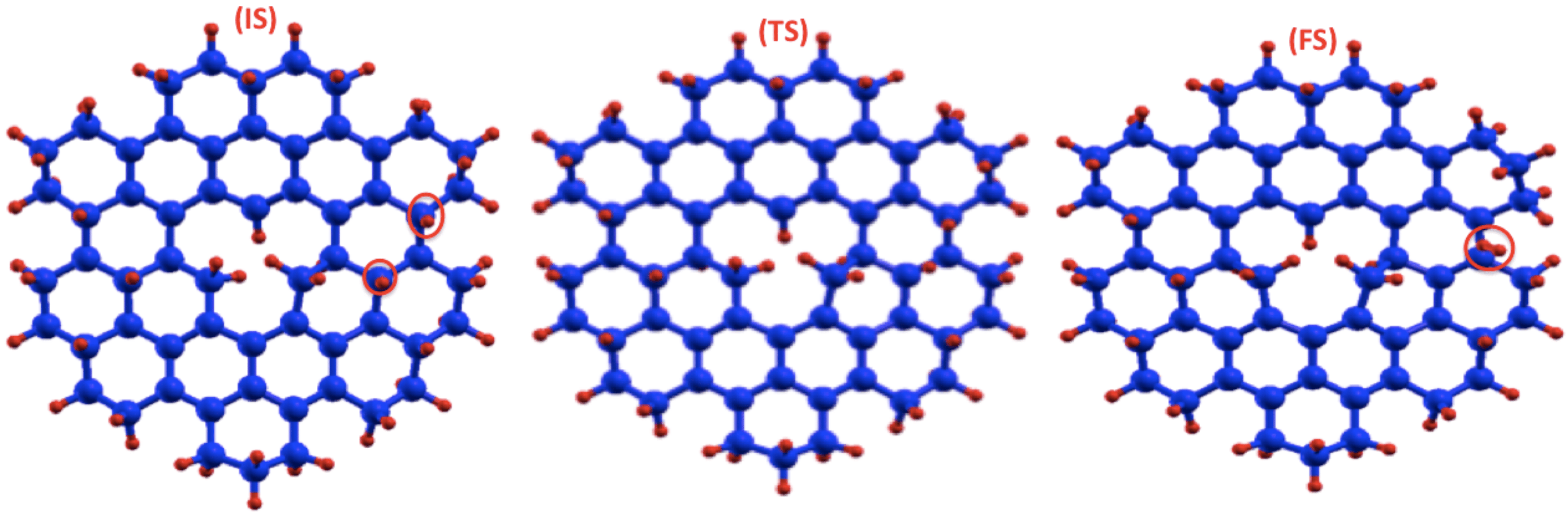}
\caption{Structures appearing along the pathway of  Figure \ref{dua}. Left: initial state (IS); center: transition state (TS); right: final structure (FS).The blue and red atoms denote carbon and hydrogen, respectively.}
\label{duaT}
\end{figure}

In larger amounts of hydrogen adsorption, VANG exhibits deformation because of the interaction with hydrogen atoms, which is identified as a buckling. This buckling depends on the adsorption of hydrogen atoms. If a hydrogen atom is adsorbed at the top of a carbon atom, the hydrogenated carbon atom buckles out of the plane directed slightly upward above the graphene plane and vice versa because of the formation of local $sp^{\mathrm{3}}$-like bondings. Here, the deformation can act as catalytically active regions to desorb hydrogen atom from the in-plane surface and dissociate absorbed molecular hydrogen into the two hydrogen atoms. The energy of the initial and final states is almost comparable, which shows good reversibility property of hydrogen uptake-release. The final state is a bit lower in energy, indicating that the reaction is an exothermic reaction; thus the adsorbed hydrogen can easily be released from the in-plane surface. This new finding solves a low reversible issue without adding any external control, which is potentially interesting for hydrogen storage applications.

Our results provide the basis for a special structure of nanographene VANG-based hydrogen storage device that relies on the specific local structure of V$\textsubscript{111}$ and its deformation to have good reversibility of hydrogen uptake-release. This novel hydrogenated nanographene VANG, exhibits excellent performance in adsorption and desorption processes. The experiment validation of this new finding must pass by synthesizing this special nanographene VANG. Thus, a novel and unique advantage of the new nanographene-VANG in exploiting its specific hydrogenated vacancy (V$\textsubscript{111}$) and its deformation surface would be realized. Here, the hydrogen uptake reaction is mediated by V$\textsubscript{111}$ of VANG. Once the hydrogen coverage is sufficiently high, the hydrogen release is possible from the in-plane surface of nanographene VANG by maintaining the structure of quintuply hydrogenated VANG (V$\textsubscript{221}$).

\section{Summary}
In this study, we obtained an effective way to store hydrogen through V$\textsubscript{111}$ of VANG, resulting in a quintuply hydrogenated vacancy (V$\textsubscript{221}$) with an activation barrier of 1.25 eV, where the initial and final states showed almost comparable energy value, indicating convenient uptake and release of hydrogen.  Once V$\textsubscript{221}$ was formed, the migration of hydrogen atoms from V$\textsubscript{221}$ to the in-plane and armchair edges is possible with an energy barrier less than 1.25 eV, resulting in the recovery of V$\textsubscript{111}$, which becomes an active site for successive hydrogen uptake through V$\textsubscript{111}$. In this special nanographene, a slightly deformed structure enhancing migration processes of hydrogen atoms from vacancy to the in-plane and armchair edges of VANG.

In larger amounts of hydrogen coverage, hydrogen is stored at vacancy as V$\textsubscript{221}$ and some hydrogen stored at the in-plane and edges of VANG. The hydrogen release from the in-plane surface shows a slightly non-endothermic reaction with a small energy barrier of less than 2 eV, indicating conveniently released hydrogen, which will be key in the application of VANG nanographene as hydrogen storage materials.

Therefore, VANG nanographene with low and high-hydrogen density has characteristics of self-catalyst reactions, leading to a small energy barrier. It exhibits a slightly exothermic release where the structure only consists of carbon and hydrogen without any additional metal catalyst, overcoming limitations of existing materials with non-exothermic hydrogen release. This study provides a basic understanding on how hydrogen is stored, migrated and released for hydrogen storage application in nanographene VANG, especially on the non-endothermic release to solve current reversibility issues. 

\section*{Acknowledgments}
The calculations were performed at the computer centers of Kyushu University. G.K.S. gratefully acknowledges fellowship support from the Japan Society for the Promotion of Science (JSPS) with numbers L21547. Y. W. and H. H. gratefully acknowledge the fellowship support from the JSPS (Grant No. 20J22909  and 21J22520, respectively).
\bibliographystyle{elsarticle-num}
\bibliography{paper.bib}

\end{document}